\documentclass[conference]{IEEEtran}

\ifCLASSINFOpdf
  \usepackage[pdftex]{graphicx}
\fi

\usepackage{paralist}
\usepackage{makecell}

\begin{document}
\title{Performance Evaluation of  Microservices Architectures using Containers}

\author{
\IEEEauthorblockN{Marcelo Amaral, Jord\`{a} Polo, David Carrera}
 \IEEEauthorblockA{Technical University of Catalonia (UPC) \\
 Barcelona Supercomputing Center (BSC)}\\
\IEEEauthorblockA{Email: marcelo.amaral, jorda.polo, david.carrera@bsc.es }
 \and
 \IEEEauthorblockN{Iqbal Mohomed, Merve Unuvar, \\
  Malgorzata Steinder}
 \IEEEauthorblockA{IBM T.J. Watson Research Center, Yorktown Heights, NY}\\
\IEEEauthorblockA{Email: iqbal, munuvar, steinder@us.ibm.com}
}

\maketitle
\thispagestyle{plain}
\pagestyle{plain}

\begin{abstract}
Microservices architecture has started a new trend for application development
for a number of reasons: (1) to reduce complexity by using tiny services; (2)
to scale, remove and deploy parts of the system easily; (3) to improve
flexibility to use different frameworks and tools; (4) to increase the overall
scalability; and (5) to improve the resilience of the system.
Containers have empowered the usage of microservices architectures by being
lightweight, providing fast start-up times, and having a low overhead.
Containers can be used to develop applications based on monolithic
architectures where the whole system runs inside a single container or inside
a microservices architecture where one or few processes run inside the
containers. Two models can be used to implement a microservices architecture
using containers: master-slave, or nested-container. The goal of this work is
to compare the performance of CPU and network running benchmarks in 
the two aforementioned models of microservices
architecture hence provide a benchmark analysis guidance for system designers.
\end{abstract}

\IEEEpeerreviewmaketitle

\section{Introduction}
\label{sec:introduction}

Virtual Machines are a widely used building block 
of workload management and deployment. They are heavily used
in both traditional data center environments and clouds (private, public and hybrid clouds). The commonly used term Virtual Machine (VM) refers to server virtualization, which can be accomplished via full virtualization or paravirtualization. In recent months, there has been a resurgence of interest in container technology, which provides a more lightweight mechanism - operating system level virtualization. Containers are lightweight and fast - a single x86 server can reasonably have 100s of containers running (memory usually ends up being the scarce resource); moreover, containers start up very quickly - under 1 to 2 seconds in most cases. There are many reasons for this resurgence but from a technical perspective, two of the biggest reasons are (i) the improvements in namespace support in the Linux kernel are available in popular distributions, and (ii) a specific implementation of containers - Docker - has successfully created an attractive packaging format, useful tools and diverse ecosystem.

Containers are operating-system-level virtualization under kernel Linux that
can isolate and control resources for a set of processes. 
Because of container does not emulate a full virtualization of the physical hardware as VM does, it
is lightweight with less overhead. 
The core of containers rely on Linux
namespace~\cite{namespace} and cGroups~\cite{cgroups}. The former is an
abstraction that wraps a set of processes appearing that they are isolated
instance. 
Linux namespace isolates the set of filesystem mount points seen by
the group of processes. cGroups organize the processes in a hierarchy tree; they
also limit, police and account the resource usage of process group. One can
run a single application within a container whose namespaces are isolated
from other processes on the system. 
Notwithstanding, the main capability of
container is to allow to run a complete copy of the Linux OS within it without
the overhead of running hypervisor. Although the kernel is shared, they have
limited access to the modules and drivers to the ones that it has leaded.
Despite a container has limited access, it can have full access to the host
when created as ``privileged". Such a privileged container might run also
another daemon inside to create other containers as a nested-container
approach; but can emerge security problems when sharing the infrastructure
with other users. Currently, there exist many container distributions such as
OpenVZ~\cite{OpenVZ}, and Docker~\cite{Docker}.

At a high-level, current Docker usage can be categorized into two classes: (i) Docker container as a lightweight server, and (ii) one process (or few related processes) per Docker container. There have been extensive studies on (i), such as \cite{Felter_2014}; in this paper, we concern ourselves with (ii). To our knowledge, this approach (related processes per container) does not have a widely used name even though the technique itself is common. For purposes of this paper, we will call it Related Processes Per Container or RPPC. RPPC is sensible from the perspective of deployment, and is also a useful building block in microservices architectures. We explain both in turn.

Consider a traditional application server that implements some business logic and talks to a remote database server. Along with the core application server, one would install auxiliary software or sidecars that provide facilities such as logging, performance monitoring, configuration management, proxying, and so on. Rather than packaging all of this software into a single unit, one can group related processes in containers and then deploy the ensemble. When some functionality needs to be updated, one need only deploy a subset of the original containers. Thus, the RPPC approach typically speeds up deployment, reduces disruption and generally empowers the devops team. It should be noted that this concept is not new to even containers - configuration management tools such as Puppet and Chef have provided this capability for years - only update what needs to be changed. What's different is that individual containers are the unit of deployment.

The RPPC concept is also useful as a building block for microservices. Microservices is a new trend in architecting large software systems wherein a system is designed as a set (dozens or even hundreds) of microservices. Microservices can be developed, managed and scaled independently. There is typically some kind of routing fabric that gets requests to a specific instance of a microservices; this routing fabric often provides load-balancing and can isolate microservices that are in a failed state. One system that provides these capabilities in a cloud or clustered environment is Google's Kubernetes. In Kubernetes, a pod is a group of containers that is a deployable unit - moreover, all containers of a pod share the same fate. While this doesn't always make sense (e.g. in the application server example above), it does make a lot of sense in microservices and especially with containers (destroying a faulty pod does no harm since the routing fabric will route to healthy one; also, since starting up containers is a lot faster than booting Virtual Machines, pod startup can be very fast).

We investigated two distinct ways to implement the RPPC concept. In the first approach, all child containers are peers of each other and a parent container (which serves to manage them). We call this implementation as master-slave throughout the paper. The second approach, where we refer as nested-container, involves the parent container being a privileged container and the child containers being inside its namespace. Our purpose is to understand the performance differences of the two approaches hence system designers can benefit from our analysis in the future.

\section{Microservices Architecture using containers}
\label{sec:microservices}
\subsection{Containers}
Containers are a mechanism that provide operating system level virtualization,
in that they can isolate and control resources for a set of processes.
Because a container does not emulate the physical hardware as a virtual 
machine does, it is lightweight with less overhead. While the concept of 
operating system level virtualization is not new (e.g. chroot/jails in BSD), 
there has been a great deal of industry interest in Linux containers and Docker 
Inc's implementation in particular.
The core of containers rely on Linux
namespaces~\cite{namespace} and cGroups~\cite{cgroups}.
Linux namespaces isolate the set of filesystem mount points seen by
a group of processes. cGroups organize the processes in a hierarchy tree; it
also limits, polices and accounts for the resource usage of process groups. One can
run a single application within a container whose namespaces are isolated
from other processes on the system.
On the other hand, one can use a container as a lightweight Linux server, by 
setting up facilities such as a process control system (e.g. supervisord) and 
and ssh server. Currently, there exist many container implementations such as
OpenVZ~\cite{OpenVZ}, Rocket~\cite{rocket} and Docker~\cite{Docker}. In this 
paper, we primarily focus on Docker containers.

Normally, a Docker container has limited access to resource on the host, but 
full access can be provided by creating it as a "privileged" container. Such a 
privileged container can access host devices and also run another Docker daemon 
inside itself to create child containers (we call this the nested-container 
approach). One must be careful with this approach as security challenges may 
emerge when sharing system resources with other users.


\subsection{Key virtual networking differences}

In recent years, there have been significant advances
in virtual networking features in Linux. Some notable mechanisms
include network namespaces, veth pairs, tap devices as well as virtual
switches such as OpenvSwitch and Linux Bridges. The mechanisms can be
used to provide a high degree of flexibility and control of networking, and form 
the basis of SDN (software defined networking) technologies such as Openstack 
Neutron. For instance, a physical host might have just one physical network 
interface, while a number of guest containers can run in isolation by having 
their own network namespaces, with tap devices wired into an OpenvSwitch. 
Moreover, one could setup tunnels between the OpenvSwitch instances on different 
machines, and can enable communication between instances. In practice, there are 
a great many ways of setting up virtual networking, with varying effects on 
throughput, latency and CPU utilization.

In this paper, we mainly focus our attention to communication that takes place
on a single host. Even in this restricted case, there are many options to 
consider. Figure 2 shows 4 different communication setups that we explore later 
in the paper, along with the various layers that must be traversed. Of course, 
best performance is achieved when a process runs on bare-metal and can utilize 
the native networking stack in the OS. The default setup with Docker is that 
containers are wired into a virtual switch such as a Linux Bridge or OpenvSwitch 
(which is wired into the physical NIC/OS networking stack). In the 
nested-container scenario, there is a second level of virtual switching that 
takes place inside the container. In the virtual machine case, there is 
typically a virtual switch as well as virtualized network drivers.

\subsection{Microservices}

Microservices have gained much popularity in industry in the last few years.
This architecture can be considered a refinement and simplification of
Service-oriented Architecture (SOA) \cite{erl_2005}. The key idea is that rather 
than architecting monolithic applications, one can achieve a plethora of 
benefits by creating many independent services that work together in concert. 
The benefits accrued include simpler codebases for individual services, ability 
to update and scale the services in isolation, enabling services to be written 
in different languages if desired and utilize varying middleware stacks and even 
data tiers for different services \cite{namiot_14}. There are costs to this 
approach as well \cite{lewis_microservices} such as the computational overhead 
of running an application in different processes and having to pay network 
communication costs rather than simply making function calls within a process. 
The proliferation of service processes almost requires automated mechanisms for 
deployment (sometimes called continuous delivery or continuous integration). 
Other challenges include deciding what the boundaries of different services are 
and determining when a service is too big.

\subsection{Achieving Microservices with Containers}

As previously stated, Docker containers are an excellent match for building 
microservices. They are lightweight, start very fast, and can wrap dependencies 
and vagaries of implementation inside themselves. For instance, a developer can 
start dozens of containers on a modest laptop of the day. They can go to a 
source such as DockerHub to download images of containers with pre-configured 
middleware, databases and applications. Such images can be extended 
with additional customization of the developer's choosing.

\begin{figure}[bth!]
    \center
    \includegraphics[width=0.48\columnwidth]{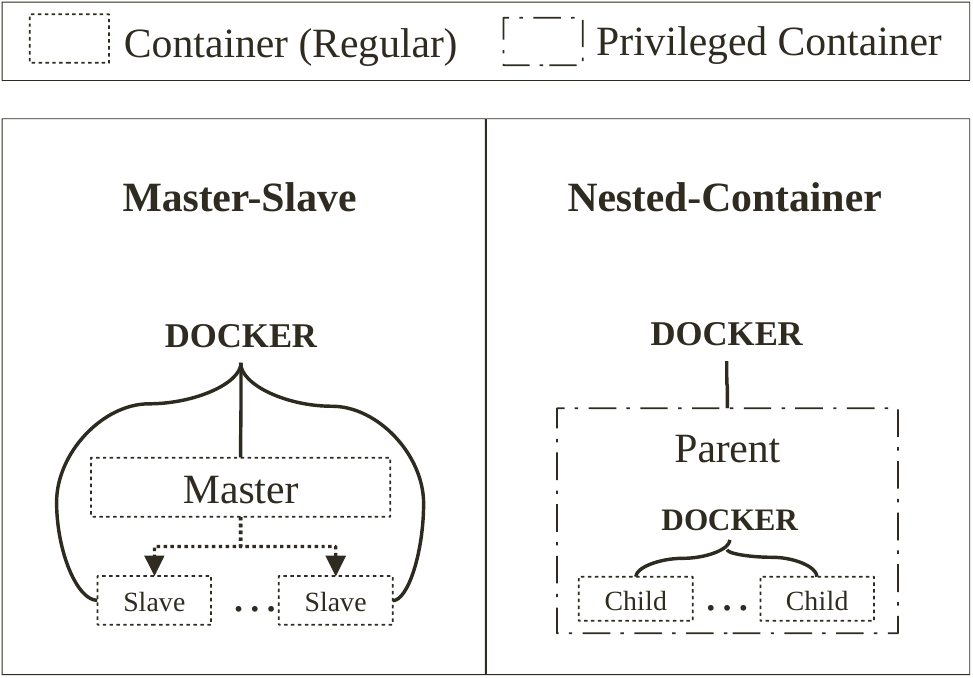}
    \caption{Overview of master-slave and nested-container models.}
    \label{fig:models}
\end{figure}

When one considers how services developed as ensembles of containers can be
deployed, two different approaches emerge: master-slave or nested-containers
as illustrated in Figure~\ref{fig:models}.

\begin{figure}[bth!]
    \center
    \includegraphics[width=0.70\columnwidth]{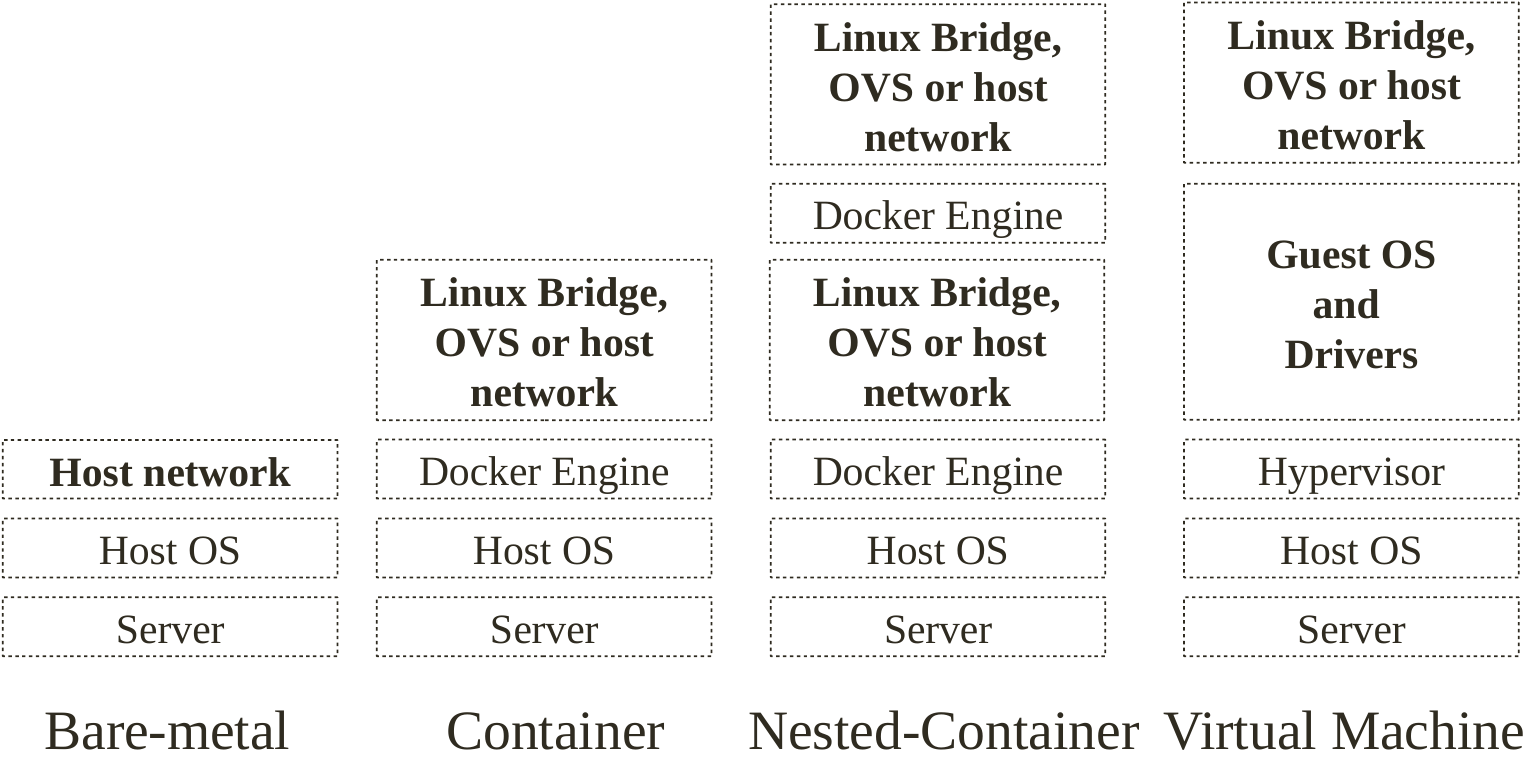}
    \caption{The network stack for bare-metal, container, nested-container and
    virtual-machine.}
    \label{fig:network_stack}
\end{figure}

\emph{The master-slave} is composed by one container as the master coordinating
other containers called slaves, in which the application process
will be running. In this approach the master needs to track the subordinates'  
containers, help their communication and guarantee that the slaves do not
interact with other containers from a different master (in this paper, in
order to 
simplify, we refer to the master-slave approach as regular-container). On the 
other hand, in the \emph{nested-container} approach, the subordinates' 
containers (the children) are hierarchically created into the main container 
(parent). The parent might be completely agnostic and just exist. The children 
run the  application process and they are limited by the parent's 
boundaries. The nested-containers approach might be easier to manage since all 
other containers are inside only one container. This approach also might benefit
from easily performing IPC, guaranteeing fate sharing, and sharing the same 
memory, disk and network. But, nested-containers approach might include more 
overhead than the master-slave approach, since it has two layers of Docker 
daemon as illustrated in Figure~\ref{fig:network_stack}.

\section{Related work}
\label{sec:related}

To the best of our knowledge, this is the first work to analyze performances
of microservices architecture using containers as nested and master-slave implementations.
However, there is an extensive work that has been done in the area of
performance comparison between virtual machines and Linux
containers~\cite{Felter_2014}. Most of the work in the literature is comparing
hypervisors to the other hypervisors or to non-virtualized
environments~\cite{Huber_2011, Hwang_2013}.

Virtual machines were first introduced by IBM on mainframes in
1970s~\cite{Creasy_1981} and since then reinvented by many companies such as
VMware, Xen and KVM. Linux containers have a long history as well. Linux-V
Server project~\cite{Soltesz_2007} was the first attempt by Linux to implement
virtual private servers. In 2007, Linux introduced native containers which
brought us today's containerization technology. 

The most popular container implementation, as of writing this paper, is Docker 
~\cite{Docker}. It is an open platform that has been used for the last couple of 
years for developers and system administrators to build, ship and run the 
applications without the VM overhead ~\cite{Docker}. Docker containers have been 
widely accepted and used by the open source cloud community and the enterprises. 
Recently, IBM started an engagement with the Docker community 
\cite{IBMContainer} and built their Beta version of container cloud 
~\cite{IBMCloud}. Companies like VMware, Warden Container, Imctfy are also 
focusing on building container clouds as the containerization technology rapidly 
gains popularity.

The significance of comparing virtualization and containerization is twofold.
One hand, virtualization helps isolation with the cost of latency and
overhead.  And, on the other hand containerization brings less overhead along
with a similar but less isolation than virtual machines. In terms of isolation
capabilities of containers, \cite{Dua_2014} studied performance of different
container implementations such as Linux, Docker, Warden Container, Imctfy and
OpenVZ along with virtual machines. As for the trade-off between performance
and isolation between containers and virtual machines, \cite{xavier_2013}
examined number of experiments for High Performance Computing environments.
Performance of network virtualization for containers is also well studied in
the Linux literature \cite{casoni_2013} however not in the nested, master-slave container,
virtual machines or bare-metal level.

Nested-containers are inspired from the ``pod" concept that is implemented by
Google for better managing Docker containers~\cite{google}. {\it Pods} are the 
smallest deployable group of containers that are placed on the same host. 
However, the performance of pods over virtual machines or bare-metal instances 
has not been studied. The contribution of our paper is to provide benchmark 
analysis for container virtualization via implementing nested and master-slave 
containers hence comparing the performance against virtual machines and 
bare-metal.

Microservices management gained popularity over the last years. Google Cloud Platform implemented and open sourced a pre-production Beta cluster management project -Kubernetes- that can be used for better management of large scale microservices ~\cite{kubernetes}. Kubernetes provides lightweight, simple and self-healing services management. With its high scalability and simplicity properties, Kubernetes promotes container technology via microservices architecture. Mesos ~\cite{mesosWeb} is another open source project that is intended to be an operating system for a datacenter. Mesos is built similar to the Linux kernel with a different level of abstraction. Mesos is composed of master(s), slave(s) and framework(s). Frameworks are designed to manage the different type of workloads hence Mesos provide a hierarchical resource management solution that increases the scalability and reduces the latency due to resource scheduling. Mesos is highly scalable (10,000s of nodes). This makes Mesos a strong open source resource management tool for microservices management. ~\cite{mesos} studied the performance of Mesos on multiple types of cluster computing such as Hadoop and MPI however did not consider the container technology. 

\section{Evaluation}
\label{sec:evaluation}

In order to explore the performance of the two container-based environments
discussed in Section~\ref{sec:microservices}, we execute five different kinds
of experiments mostly focused on CPU and network performance. In Experiment 1
(Section~\ref{sub:exp1}), we consider the execution of a CPU-intensive
benchmark in order to verify whether there is any performance difference between
the studied approaches. In Experiment 2 (Section~\ref{sub:exp2}), we evaluate
the overhead of virtual container creation that might be used for management
decisions. In Experiment 3 (Section~\ref{sub:exp3}), we evaluate the overhead
of creating nested-containers. In Experiment 4 (Section~\ref{sub:exp4}), we compare the proposed approaches, but with a focus on network performance
of local traffic on one host. Finally, in Experiment 5
(Section~\ref{sub:exp5}), we also evaluate the network, but with a focus on
remote traffic across two hosts.

The main goal of these experiments is to study the performance and overhead of
nested-containers, which might play a key role in the implementation of
microservices architectures. In all experiments we compare the performance of
the following environments:
\begin{inparaenum}[\itshape i\upshape)]
 \item bare-metal
 \item regular containers (representing the master-slave approach)
 \item nested-containers
 \item virtual machines.
\end{inparaenum}

\subsection{Evaluation infrastructure}

The machines that are used to run the experiments are several 2-way Intel Xeon 
E5-2630L, each one composed by 2 sockets, 6 cores per socket, 2 hyper-threads 
per core at 2GHz, 64GB of RAM and 1 Gbps Network Interface Card (NIC). They are 
Linux box running Ubuntu 14.04 (Trusty Tahr) with Linux Kernel 3.13. Those 
machines are in the same rack and are connected with 2 stacked Gigabit Cisco 
Switch model 3750X with 48port each, connected through StackWise+ connector. For 
containers we used Docker 1.0.1 build 990021a, while virtualization was provided 
by KVM 2.0.0 configured with Intel Virtualization Technology (VT-x) and network 
Gigabit mode (virtio).
Experiments focused on CPU (\ref{sub:exp2} and~\ref{sub:exp3}) are based on the 
Sysbench benchmark~\cite{sysbench} version 0.4.12.  Experiments focused on 
network (\ref{sub:exp4} and~\ref{sub:exp5}) use Netperf version 2.6.0, and the 
machines are configured with Open vSwitch version 2.0.2, and the Linux Bridge
version natively available in Ubuntu 14.04. All the programs were compiled using 
gcc version 4.8.2 and Python 2.7.6.

\subsection{Experiment 1: CPU Performance Evaluation}
\label{sub:exp1}

The goal of this first experiment is to compare the computing performance of
different kinds of environments: bare-metal, regular containers,  
nested-containers, and virtual machines. In order to evaluate the performance, 
we selected the Sysbench benchmark~\cite{sysbench} running in CPU mode, where
each request calculates prime numbers up to a certain value specified by the
\texttt{cpu-max-primes} option, in this experiment set to 40,000; all
calculations are performed using 64-bit integers.

This experiment measures the mean execution time of running Sysbench,
increasing the number of concurrent Sysbench instances, from 1 up to 64. For
bare-metal, we simply run multiple Sysbench instances on the host machine. For
containers, we execute multiple containers, each one running a single Sysbench
instance. And for virtual machines, we also run multiple virtual machines
executing a single Sysbench instance. No resource constraints are set of
containers or virtual machines, so the scalability is expected to grow
linearly with the number of available CPU cores.

\begin{figure}[bth!]
    \center
    \includegraphics[width=0.75\columnwidth]{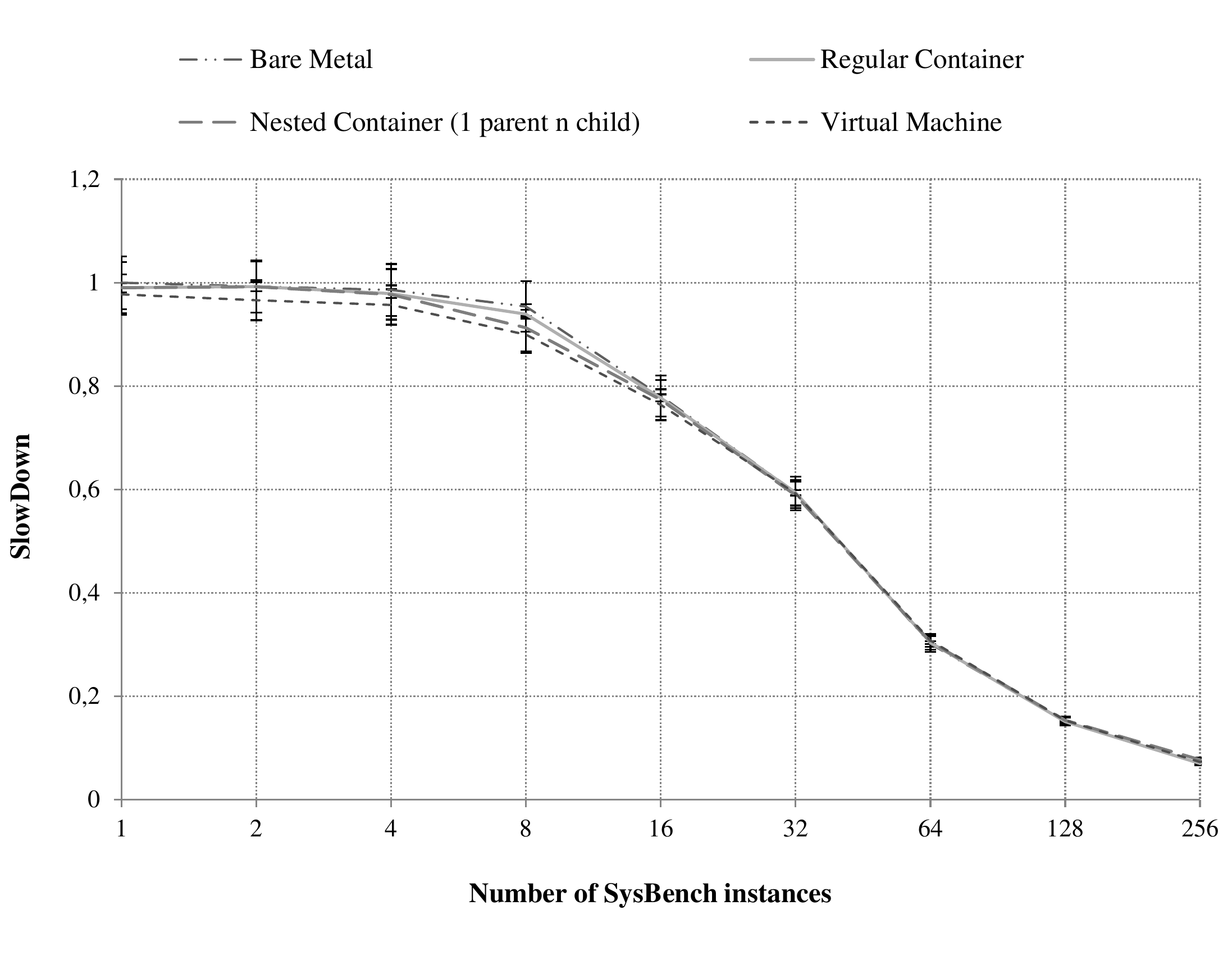}
    \caption{Observed slowdown of Sysbench with increasing number of instances
    relative to running a single Sysbench instance in bare-metal}
    \label{fig:slowdown}
\end{figure}

The results of this experiment are illustrated in Figure~\ref{fig:slowdown},
which shows the observed slowdown of running Sysbench when increasing the
number of instances. In particular, the Figure shows the average of 10
executions for each one of the tested environments: bare-metal, regular
containers, nested-containers, and virtual machines. The baseline to measure
the slowdown is the execution time of one single Sysbench instance running on
bare-metal.

As it can be observed in Figure~\ref{fig:slowdown}, all environments display a
similar behavior, confirming there is no significant performance impact for
CPU-intensive executions when running on containers or virtual machines
compared to bare-metal. Containers run natively in the operating system, and
they are only isolated by a lightweight layer (cgroups and namespaces), so as
expected they basically perform as well as bare-metal.  But even virtual
machines perform as much thanks to improved virtualization support in modern
processors.  It should also be noted that slowdown degrades slowly up to 8-16
concurrent instances, and significantly faster after that. This is basically
related to the characteristics of the experimental machine, which has 12
cores.  When more than 12 instances are running, they have to share the same
cores, leading to increased context switching.

\subsection{Experiment 2: Comparing Overhead of Virtual Container Creation}
\label{sub:exp2}

While there is no significant performance impact for CPU bound applications
under different environments, there may be a higher variation when considering
the management of the proposed approaches.

This experiment evaluates the performance impact of creating the
\textit{hosting entities} for different virtualization technologies:
containers, nested-containers (Microservices), and virtual machines in the
context of server virtualization. The goal of this experiment is to evaluate
the scalability of managing different kinds of virtual containers.

We measure the time to create an increasing number concurrent \textit{hosting
entities} (containers or virtual machines) for each one of the proposed
approaches, from 1 up to 64. Each \textit{hosting entity} simply launches a
dummy application that takes a negligible amount of time (in particular, we
used Sysbench as in Experiment 1, this time configured with
\texttt{cpu-max-primes} set to 1), effectively allowing us to compare creation
times under different environments. For regular containers, we measure elapsed
time between container start-up and exit.  For nested-containers, we measure
the elapsed time between starting-up and exiting the parent container, which
also includes loading a locally-stored child image as well as starting-up and
exiting a single child container.  Finally, for a virtual machine, we measure
the time to create a virtual machine domain, start the domain and delete the
domain.

\begin{figure}[bth!]
    \center
    \includegraphics[width=0.75\columnwidth]{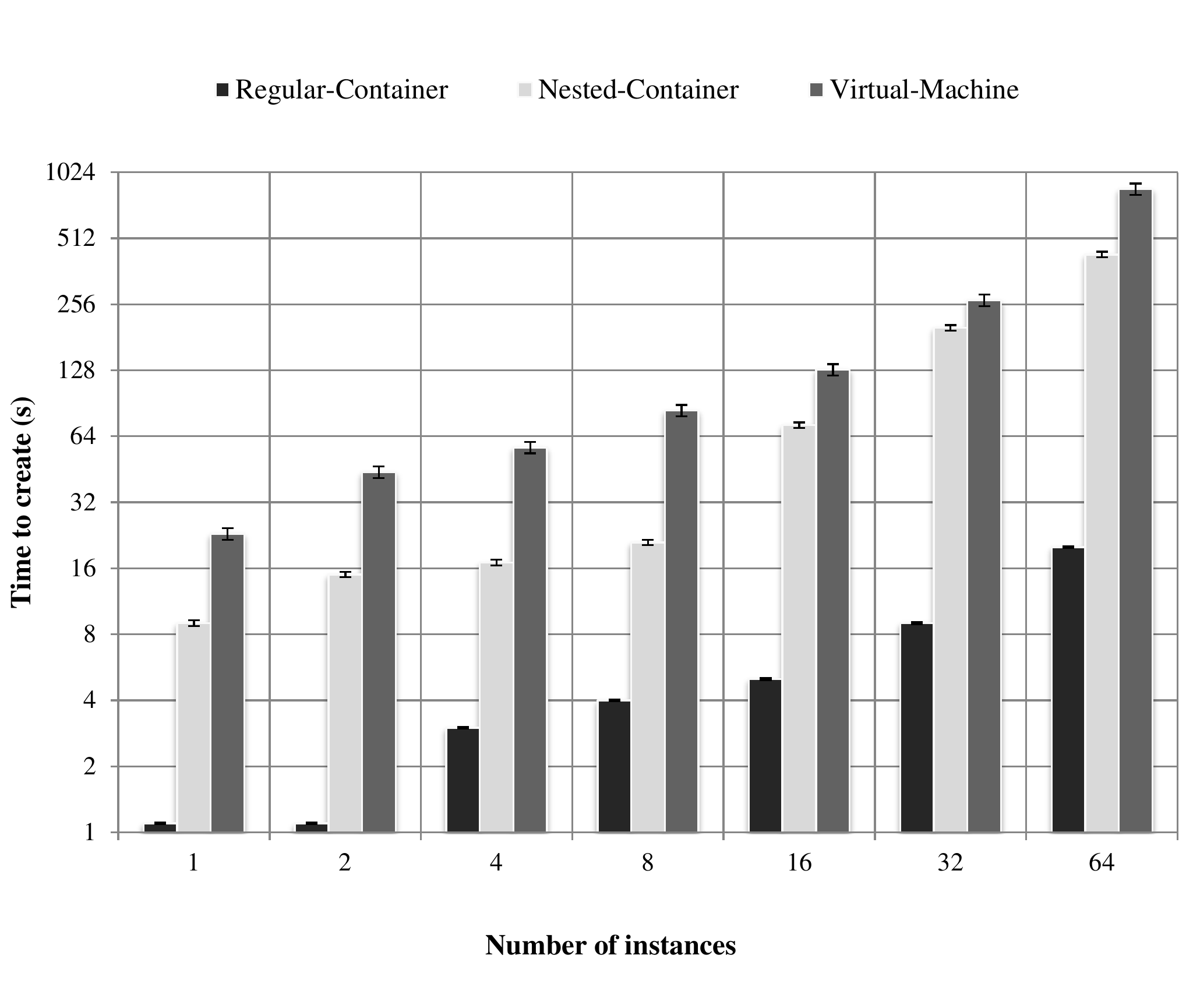}
    \caption{Time to create an increasing number of instances of virtual
    containers (base 2 log scale in both axes). Where the nested-container is a 
fully initialized parent plus one child.}
    \label{fig:creation-time}
\end{figure}

The results of our experiments are available in
Figure~\ref{fig:creation-time}, which shows the measured time to create
different number of instances under each environment.  As expected regular
containers are always the fastest approach, followed by nested-containers and
virtual machines.
While the creation of a single nested-container has almost 8 times more
overhead than the creation of one regular container, the creation of  
nested-containers is still more than twice as fast as virtual machines.  

This additional overhead for nested-containers is related to the initialization 
of Docker in the parent container, which also involves loading an image stored 
locally on the host and the creation of the child container itself. The 
main overhead is related to the image loading, which takes in average 6.2s. In 
order to avoid this loading time, a parent container can be created with 
a shared pre-created (read-only) volume already contain the child image loaded. 
We verified this approach creating a parent plus a child container with the 
parent sharing a preloaded volume, and the creation time dropped from 8s to 
1.7s. However, such approach has some drawbacks since many parents will be 
sharing the same volume, such as concurrency and security problems (especially 
the last one, since the parent is a privileged container). Because of this 
trade-off, our focus in this paper is to show the overhead to fully initialize 
a parent container.

Also, note that when creating more than 8 nested-containers, the overall
creation time seems to increase more than linearly, and becomes a lot closer
to virtual machines than regular containers. The host machine only has 12
cores, so the behavior when overloading the cores is significantly different.
However, nested-containers are still twice as fast as virtual machines in most
scenarios.


\subsection{Experiment 3: Overhead of Nested-Container Creation}
\label{sub:exp3}

\begin{table*}
\renewcommand{\arraystretch}{1.3}
\centering
\caption {Time to create nested-containers with different ratios of parent
    to children containers}
\begin{tabular}{|l||c|c|c|c|c|c|c|c|c|c|}\hline
\diaghead{\theadfont Parenttttttttt Child}%
{\\ \# Child }{\# Parent}&1&  2                &	4          &	8                 &	16                &	32             &	64            &	128             &	256\\    \hline
\hline
1       &9  s $\sigma$=1.0&15 s $\sigma$=0.5 &17 s $\sigma$=0.5 &21 s $\sigma$=0.4 &72 s $\sigma$=1.3   &200 s $\sigma$=0.3 &432 s $\sigma$=10.5 &1475 s $\sigma$=146.5 &2313 s $\sigma$=160.8\\    \hline
2       &9  s $\sigma$=0.5&10 s $\sigma$=0.51&12 s $\sigma$=0.43&25 s $\sigma$=2.0 &96 s $\sigma$=4.0   &239 s $\sigma$=2.5 &475 s $\sigma$=14.5	&1518 s $\sigma$=163.0 &-	\\   \hline
4       &10 s $\sigma$=0.5&12 s $\sigma$=0.5 &17 s $\sigma$=0.5 &37 s $\sigma$=2.0 &113 s $\sigma$=8.0  &255 s $\sigma$=0.6 &514 s $\sigma$=12.0 &-		 &-  \\    \hline
8       &12 s $\sigma$=0.5&14 s $\sigma$=0.53&22 s $\sigma$=3.56&64 s $\sigma$=1.5 &131 s $\sigma$=6.1 &278 s $\sigma$=7.0 &-		&-	     &-  \\    \hline
16      &16 s $\sigma$=1.0&22 s $\sigma$=1.0 &35 s $\sigma$=1.5&66 s $\sigma$=0.5   &139 s $\sigma$=1.5&-	    &-	    &-		 &-  \\    \hline
32      &27 s $\sigma$=1.5&40 s $\sigma$=0.5 &47 s $\sigma$=4.81&92 s $\sigma$=3.5  & -		           &-	       &-	    &-	     &-  \\    \hline
64      &46 s $\sigma$=2.0&46 s $\sigma$=2.0 &69 s $\sigma$=2.02&-	                 &-		            &-	       &-	    &-	     &- \\    \hline
128     &65 s $\sigma$=3.0&77 s $\sigma$=1.0 &-      	          &-	             &-		            &-	    &-	       &-	     &-  \\    \hline
256     &111 s $\sigma$=3.5&-	            &-	                   &-	              &-		          &-	    &-	       &-	     &-   \\    \hline
\end{tabular}
\label{tab:creation-time-nested}
\end{table*}

\begin{figure}[bth!]
    \center
    \includegraphics[width=0.80\columnwidth]{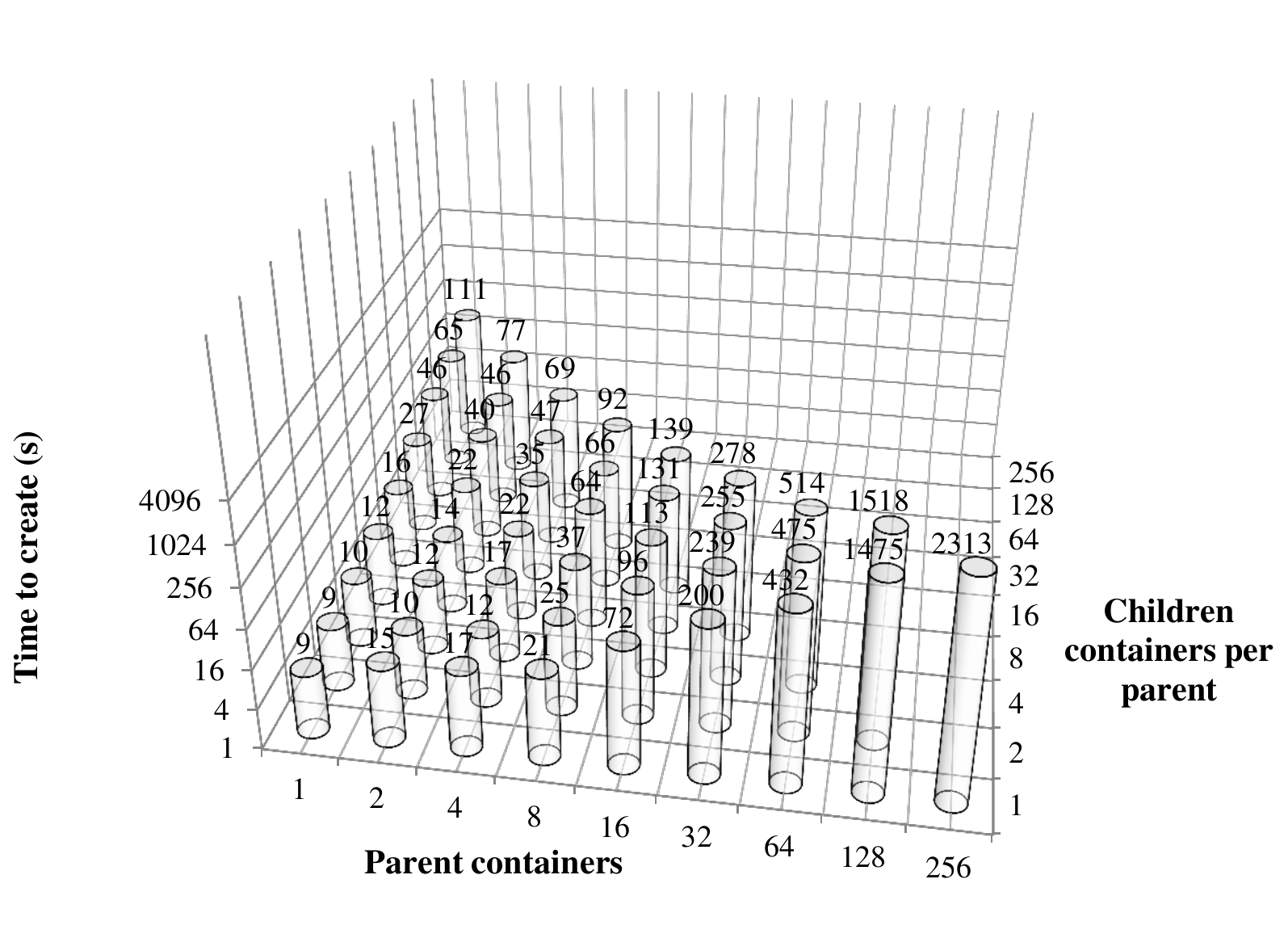}
    \caption{Time to create nested-containers with different ratios of 
fully initialized parent and child containers (base 2 log scale)}
    \label{fig:creation-time-nested}
\end{figure}

Following Experiment 2, which evaluates the creation of parent containers with
a single child container, this Section focuses on the overhead of creating
parents with multiple children.


The goal of this experiment is to measure the overhead of different ratios of
parent to child containers.  In each execution, we change the number of
concurrent parent containers, from 1 up to 256, also varying the number of
concurrent children in each parent, again from 1 up to 256. The maximum number
of concurrent children running at a time is 256. We measure the elapsed time,
from start-up to exiting the parent container, which involves starting the
Docker daemon, loading a locally-stored image, starting-up all children
containers, and exiting all of them.

The results of this experiment are summarized in
Figure~\ref{fig:creation-time-nested} which shows the time to create  
nested-containers with different configurations of parent to child containers 
(for clarity, we also include the same average times as well as the standard 
deviation in Table~\ref{tab:creation-time-nested}). As it can be observed, the 
creation of only one parent with 256 children takes 111 seconds to complete. 
However, 256 parents with only one child takes 2312 seconds. This is due to the
bottleneck's relation to the initialization of the parent container, which is
supposed to initialize Docker and load children image. Once the parent is 
running, the creation of child containers does not take more than the creation 
of a regular container, and will be faster than the creation of new parents.
Moreover, the results show that with less than 16 parents, the creation of
256 child containers benefits from the concurrent child's creation across
different parents. The creation of 256 children into a single parent takes 111
seconds, while the creation of the same number of children across 4 parents
only takes 69 seconds. When we are overloading the host machine with more 
parents than available cores, the creation time is even higher (more than 139 
seconds).

\subsection{Experiment 4: Network Performance - Local traffic in one host}
\label{sub:exp4}

As described in Section~\ref{sec:microservices}, microservices architectures
may involve multiple applications communicating with the same machine. The goal
of this experiment is to measure network performance, with a focus on studying
the communication overhead of different technologies when running on the same
host.

In order to evaluate the network performance, we select the Netperf
benchmark~\cite{netperf}. In particular, the Netperf tests used to evaluate
performance are TCP Stream (\texttt{TCP\_STREAM}) and TCP Request$/$Response
(\texttt{TCP\_RR}). \texttt{TCP\_STREAM} is a simple test that transfers a
certain amount of data from a client running \texttt{netperf} to a server
running \texttt{netserver}. This test calculates the throughput and does not
include the time to establish the connection. On the other hand,
\texttt{TCP\_RR} is a synchronous test that consists of exchanging requests
and responses (transactions), and which can be used to infer one-way and
round-trip latencies.  In particular, \texttt{TCP\_RR} executions were
configured to run in burst mode in order to have more than one transaction at
the same time, and the socket buffer size of connection data set to 256K.
Throughput and round trip latency per transaction were measured with Netperf
under different environments: bare-metal, containers, and virtual machines;
and also under different network virtualization technologies, such as
Host-Network, OpenvSwitch, and Linux Bridge.

In this experiment, both server and client (\texttt{netserver} and
\texttt{netperf} respectively) were executed in different configurations: host
to host, container to host, host to container, virtual machine to host, host
to virtual machine.  Additionally, each environment can be configured with
different network virtualization technologies, such as Linux Bridge or
OpenvSwitch~\cite{OVS}. Containers might also be configured as Host-Network
where there is no virtualization layer and containers use the native host
network stack directly. While Host-Network might provide certain performance
improvement, it also has security implications since it doesn't provide any
network isolation. In this experiment, OpenvSwitch is configured only for
routing packets and there is no additional encapsulation, while Linux Bridge
is combined with forwarding NAT iptable rules. Since Docker is not yet fully
integrated with OpenvSwitch, it requires additional configuration, which
involves creating a virtual interface (veth pairs), binding one in the 
OpenvSwitch bridge and another one in an already started container.

\begin{table*}[!t]
\renewcommand{\arraystretch}{1.3}
\centering
\caption{Network throughput and latency evaluation for different
    configurations of client/server under bare-metal, container
and virtual machine on a single host machine}
\begin{tabular}{|l|c|c|c||c|c|c|}
\hline
& \multicolumn{3}{c||}{Throughput} & \multicolumn{3}{c|}{Latency}                                                                \\
\hline
(Client - Server) & \multicolumn{1}{l|}{Host-Network} & \multicolumn{1}{l|}{Linux Bridge } & \multicolumn{1}{l||}{Open vSwitch} & \multicolumn{1}{l|}{Host-Network} & \multicolumn{1}{l|}{Linux Bridge } & \multicolumn{1}{l|}{Open vSwitch}  \\
\hline
\hline
Host - Host            & 35.71 Gbps $\sigma$=0.32 & -                        & -                        & 102.77 $\mu s$	$\sigma$=0.95 & -                               & -                               \\ \hline
Container - Host       & 35.13 Gbps $\sigma$=0.48 & 15.82 Gbps $\sigma$=0.36 & 16.01 Gbps $\sigma$=0.47 & 104.48$\mu s$ $\sigma$=1.45 & 231.97$\mu s$ $\sigma$=5.3   & 229.37$\mu s$	$\sigma$=6.38 \\ \hline
Host - Container       & 34.96 Gbps $\sigma$=0.63 & 15.96 Gbps $\sigma$=0.51 & 16.86 Gbps $\sigma$=0.35 & 105.0$\mu s$	$\sigma$=1.94  & 230.17$\mu s$ $\sigma$=7.35  & 217.76$\mu s$	$\sigma$=4.63 \\ \hline
Virtual machine - Host & - 						  & 8.64 Gbps $\sigma$=0.28  & 7.94 Gbps $\sigma$=0.69         & - 	                       & 424.92$\mu s$ $\sigma$=14.09 & 465.53$\mu s$ $\sigma$=43.57             \\ \hline
Host - Virtual machine & - 						  & 9.24 Gbps $\sigma$=0.27  & 8.77 Gbps $\sigma$=0.55         & - 	                       & 397.53$\mu s$ $\sigma$=12.08 & 420.14$\mu s$ $\sigma$=27.09            \\ \hline
\end{tabular}
\label{tab:net_comp_table1-localonly}
\end{table*}

The results of this experiment are presented in
Table~\ref{tab:net_comp_table1-localonly}. All executions were repeated
5 times, and the table includes average time and standard deviation. In this
experiment, it should also be noted that, even though the host is composed of a
single NIC of 1 Gigabit, the throughput is higher than 1Gbps. 
Since the client and the server are running on the same host, the 
network packets are not going through the NIC, they are going to 
the loopback interface. The bottleneck is the CPU throughput.

As it can be observed in Table~\ref{tab:net_comp_table1-localonly}, the
network performance when using containers is generally higher than that of
virtual machines, and it can be as fast as bare-metal under certain
configurations. For instance, when the containers are configured with
Host-Network, they basically display the same performance as bare-metal in
terms of both, throughput and latency. On the other hand, when containers or
virtual machines are configured with Linux Bridge or OpenvSwitch, there is a
significant performance impact.  Even though OpenvSwitch is supposed to
achieve higher performance than Linux Bridge, as stated in
\cite{callegati_2014}, our results show that their behavior in terms of
throughput and latency under these configurations is similar, and their
performance is not significantly different, only achieving approximately half
of the throughput, and almost twice as much latency.  Finally, it should also
be noted that even if virtual machines are accelerated to provide high
computing performance, the network still lacks behind in terms of performance
when compared to containers, and it's approximately twice as slow, even when
compared against the same network virtualization technologies.



\begin{table}[h]
\renewcommand{\arraystretch}{1.3}
\centering
\caption {Network throughput and latency evaluation of nested-containers}
\begin{tabular}{|c|c|c|c|}
\hline
\begin{tabular}[c]{@{}c@{}}
 Parent \\ (privileged)
\end{tabular} & Child         & Throughput & Latency \\ \hline
\hline
 Host-network & Host-network  & 33.74 Gbps $\sigma$=2.3 & 109.31$\mu s$ $\sigma$=7.66\\ \hline
 Host-network & Linux Bridge  & 16.06 Gbps $\sigma$=0.73 & 228.90$\mu s$ $\sigma$=10.52\\ \hline
 Linux Bridge & Host-network  & 15.56 Gbps $\sigma$=0.97 & 236.72$\mu s$ $\sigma$=14.72\\ \hline
 Linux Bridge & Linux Bridge  & 12.53 Gbps $\sigma$=0.75 & 293.84$\mu s$ $\sigma$=18.45\\ \hline
 Open vSwitch & Host-network  & 16.9 Gbps $\sigma$=0.44 & 217.25$\mu s$ $\sigma$=5.63\\ \hline
 Open vSwitch & Linux Bridge  & 12.19 Gbps $\sigma$=0.57 & 301.54$\mu s$ $\sigma$=14.63\\ \hline
\end{tabular}
\label{tab:net_comp_table2}
\end{table}

In addition to studying the performance of different virtualization
environments, in Table~\ref{tab:net_comp_table2} we also show the result of
evaluating network throughput and latency of nested-containers under different
combinations of network configurations for parent and child containers.  As
expected, when a parent or a child container is configured with Host-Network,
the performance is significantly better than that of Linux Bridge or
OpenvSwitch, which are once again approximately twice as slow in terms of
both, throughput and latency.  However, while Host-Network provides higher
performance, it also has certain security trade-offs, especially when using it
in the parent container. Parent containers are privileged, so they might be
granted access to all the packets in the host machine. Hence Host-Network is
only a viable configuration when used in the child containers, since they are
not allowed to access the host network.  Note that in this experiment
OpenvSwitch couldn't be used in the child container due to privilege-related
issues, so OpenvSwitch parent containers are only compared against
Host-Network and Linux bridge child containers.


\subsection{Experiment 5: Network Performance - Remote traffic across two hosts}
\label{sub:exp5}

This experiment is similar to the previous experiment described in section
\ref{sub:exp4}, except network performance is measured across two hosts
interconnected by a physical 1 Gigabit switch.

Netperf tests are once again \texttt{TCP\_STREAM} and \texttt{TCP\_RR} with
the same configuration.  However, in this experiment, \texttt{netserver} and
the \texttt{netperf} client runs on different machines. The server runs
directly on bare-metal in one machine and the client runs in all the studied
environments: bare-metal, containers, and virtual machines. As in the previous
experiment we also evaluate different network virtualization technologies,
including Linux Bridge, OpenvSwitch, and Host-Network (only
available for containers).

\begin{table*}
\renewcommand{\arraystretch}{1.3}
\centering
\caption{Network throughput and latency evaluation for different
    configurations of client/server under bare-metal, container
and virtual machine across two hosts}
\begin{tabular}{|l|c|c|c||c|c|c|}
\hline
& \multicolumn{3}{c||}{Throughput} & \multicolumn{3}{c|}{Latency}                                                                \\ \hline
(Client - Server) & \multicolumn{1}{l|}{Host-Network} & \multicolumn{1}{l|}{Linux Bridge } & \multicolumn{1}{l||}{Open vSwitch} & \multicolumn{1}{l|}{host-network} & \multicolumn{1}{l|}{Linux Bridge } & \multicolumn{1}{l|}{Open vSwitch}  \\ \hline
\hline
Host - Host            & 142.21 Mbps $\sigma$=8.64 & -                        & -                        & 25.97 ms	$\sigma$=1.7 & -                     & -                  \\ \hline
Container - Host       & 157.92 Mbps $\sigma$=1.06 & 154.51 Mbps $\sigma$=5.22& 157.25 Mbps $\sigma$=3.95& 23.29 ms $\sigma$=0.15& 23.83 ms $\sigma$=0.82& 23.40 ms	$\sigma$=0.58\\\hline
Virtual machine - Host & - 						   &135.92 Mbps $\sigma$=6.77 & 136.92 Mbps $\sigma$=5.37& - 	                 & 27.13 ms $\sigma$=1.31& 26.5 ms $\sigma$=1.23\\\hline
\end{tabular}
\label{tab:net_comp_table1-remote}
\end{table*}

The results of this experiment are presented in
Tables~\ref{tab:net_comp_table1-remote} and~\ref{tab:net_comp_table2-remote}.
The former is focused on comparing network performance of bare-metal,
containers, and virtual machines, while the latter compares different
configurations of nested-containers.  All executions were repeated
5 times, and the tables include average time and standard deviation for each
configuration.


As it can be observed in Table~\ref{tab:net_comp_table1-remote}, throughput is
significantly lower and latency is higher in this experiment (from \textit{Gbps}
and \textit{$\mu s$} in the previous experiment to \textit{Mbps} and
\textit{ms} here). However, one of the consequences of using the physical
network is that the bottleneck is now the physical network itself, and the
performance of different configurations of virtual containers and network
virtualization technologies is not as varied. The major difference is
latency when using virtual machines, which is approximately 12\% higher.  This
is due to virtual machine simulates the entire network stack in the guest
operating system, which involves additional overhead when compared with
containers.

\begin{table}[h]
\renewcommand{\arraystretch}{1.3}
\centering
\caption{Network throughput and latency evaluation of nested-containers
connected to a remote host}
\begin{tabular}{|c|c|c|c|}
\hline
\begin{tabular}[c]{@{}c@{}}
 Parent \\ (privileged)
\end{tabular} & Child         & Throughput & Latency \\ \hline
\hline
 Host-network & Host-network  & 148.98 Mbps $\sigma$=5.33 & 24.72 ms $\sigma$=0.92\\ \hline
 Host-network & Linux Bridge  & 147.39 Mbps $\sigma$=4.78 & 24.98 ms $\sigma$=0.83\\ \hline
 Linux Bridge & Host-network  & 147.29 Mbps $\sigma$=4.85 & 25.01 ms $\sigma$=0.81\\ \hline
 Linux Bridge & Linux Bridge  & 148.31 Mbps $\sigma$=4.79 & 24.83 ms $\sigma$=0.81\\ \hline
 Open vSwitch & Host-network  & 147.75 Mbps $\sigma$=5.75 & 24.93 ms $\sigma$=0.97\\ \hline
 Open vSwitch & Linux Bridge  & 147.86 Mbps $\sigma$=4.99 & 24.91 ms $\sigma$=0.85\\ \hline
\end{tabular}
\label{tab:net_comp_table2-remote}
\end{table}


Table~\ref{tab:net_comp_table2-remote} summarizes the evaluation of nested-containers under different combinations of network configurations for parent
and child containers. Again, once the physical network becomes the bottleneck
there is no significant performance impact on comparing different
combinations and configurations of network virtualization technologies. But as
described in Section~\ref{sub:exp4}, Host-Network still poses security
concerns when executed on the parent container.


\section{Conclusions}
\label{sec:conclusions}

Containers are gaining momentum because they offer lightweight OS
virtualization capabilities. They are commonly used to host single processes in
isolation on the system. While they offer clear advantages in terms of lightness
and performance under several circumstances, they show limitations from the
infrastructure management perspective. On the other hand, Server Virtualization
has been widely adopted across sectors and industries because it provides simple mechanisms to
manage the infrastructure, and group processes and applications. But it
introduces several significant penalties in terms of deployment time, memory
consumption and processing overheads that vary with the nature of the applications that they
host.

This paper explores the use of Related Processes Per Container (RPPC) as an
abstraction to leverage Containers technology, but overcoming their limitations
from the point of view of infrastructure management and application deployment.
RPPCs allow for the creation of ensembles that encapsulate sets of related
processes that are related and provide sets functionalities required by an
application once it is deployed. The RPPC approach typically speeds up
deployment, reduces disruption and generally empowers the devops team.
For this reason, RPPCs are particularly suitable for the implementation of
microservices.

Through the experiments of the paper, we evaluate the performance impact of
choosing between the two models for implementing RPCCs that we have presented:
in the first approach (``master-slave") all child containers are peers of each
other and a parent container which serves to manage the; 
in the second approach (``nested-container"), involves the parent
container being a privileged container and the child containers being in its 
namespace.

Our results show that the nested-containers approach is a
suitable model, thanks to improved resource sharing (same memory and disk),
easily performing IPC and guaranteeing fate sharing among the containers in
the same nested-container. The results show that nested-containers don't have
a significant impact on the performance of CPU, however, there are some
trade-offs in terms of network performance compared to bare-metal and regular
containers. In any case, they add some of the simplicity that Virtual Machines
offer in terms of infrastructure management flexibility and ease of
workload deployment.

Finally, it might be worth to consider the trade-off when selecting the most
appropriated model to implement microservices architecture relying on network
performance, security and simplicity. In our future work, we are planning to
implement an extension on Docker to fully support OVS and study the performance of a real
application implementation based on microservices architecture using
nested-containers.  Additionally, we will study the overhead related to the
control plane.

\section{Acknowledgments}
{\small This project is supported by the IBM/BSC Technology Center for Supercomputing collaboration
agreement. It has also received funding from the European Research Council (ERC) under the European Union's Horizon 2020 research and innovation programme (grant agreement No 639595). It is also partially supported by the Ministry of Economy of Spain under contracts TIN2012-34557 and 2014SGR1051 and by the BSC-CNS Severo Ochoa program (SEV-2011-00067).}

\bibliographystyle{IEEEtran}
\bibliography{bibliography}

\end{document}